\documentclass{article}
\usepackage[T1]{fontenc}
\usepackage{wrapfig}
\usepackage[portuges,english]{babel}
\usepackage{amssymb,latexsym,amsmath,color,mathrsfs,ifsym,graphics,stmaryrd} 
\usepackage[colorlinks,linkcolor=blue,urlcolor=blue,citecolor=black,
plainpages=false,pdfpagelabels,breaklinks]{hyperref}

\title{Epistemological vs. Ontological Relationalism in\\Quantum Mechanics: Relativism or Realism?}

\author{{\sc C. de Ronde}\thanks{Fellow Researcher of the Consejo
Nacional de Investigaciones Cient\'{\i}ficas y T\'ecnicas.} {\sc \ and R. Fernandez Moujan}}
\date{}

\usepackage[margin=2.5cm]{geometry}

\begin{document}
\maketitle

\begin{center}
\begin{small}
Philosophy Institute Dr. A. Korn \\ 
Buenos Aires University, CONICET - Argentina \\
Center Leo Apostel and Foundations of  the Exact Sciences\\
Brussels Free University - Belgium \\
\end{small}
\end{center}

\bigskip

\begin{abstract}
\noindent In this paper we investigate the history of relationalism and its present use in some interpretations of quantum mechanics. In the first part of this article we will provide a conceptual analysis of the relation between substantivalism, relationalism and relativism in the history of both physics and philosophy. In the second part, we will address some relational interpretations of quantum mechanics, namely, Bohr's relational approach, the modal interpretation by Kochen, the perspectival modal version by Bene and Dieks and the relational interpretation by Rovelli. We will argue that all these interpretations ground their understanding of relations in epistemological terms. By taking into account the analysis on the first part of our work, we intend to highlight the fact that there is a different possibility for understanding quantum mechanics in relational terms which has not been yet considered within the foundational literature. This possibility is to consider relations in (non-relativist) ontological terms. We will argue that such an understanding might be capable of providing a novel approach to the problem of representing what quantum mechanics is really talking about.
\medskip
\end{abstract}


\renewenvironment{enumerate}{\begin{list}{}{\rm \labelwidth 0mm
\leftmargin 0mm}} {\end{list}}

\newcommand{\ita}{\textit}
\newcommand{\mcal}{\mathcal}
\newcommand{\mfrak}{\mathfrak}
\newcommand{\mbb}{\mathbb}
\newcommand{\mrm}{\mathrm}
\newcommand{\msf}{\mathsf}
\newcommand{\mscr}{\mathscr}
\newcommand{\lra}{\leftrightarrow}
\renewenvironment{enumerate}{\begin{list}{}{\rm \labelwidth 0mm
\leftmargin 5mm}} {\end{list}}

\newtheorem{dfn}{\sc{Definition}}[section]
\newtheorem{thm}{\sc{Theorem}}[section]
\newtheorem{lem}{\sc{Lemma}}[section]
\newtheorem{cor}[thm]{\sc{Corollary}}
\newcommand{\Proof}{\textit{Proof:} \,}
\newcommand{\cqd}{{\rule{.70ex}{2ex}} \medskip}

\bigskip

\bigskip

\section*{Introduction}

In this article we attempt to discuss the possibility of providing a relational account of quantum mechanics. For such purpose we intend to clarify which are the main distinctions between substantivalism, relationalism and relativism. We will argue that, apart from the numerous interpretations which consider relationalism from an epistemological perspective, there is also the possibility to understand relations from a  (non-relativist) ontological viewpoint. The paper is organized as follows. In the first section we provide a short account of the relation between physics, philosophy and sophistry as related to realism and relativism. In section 2 we address relational and substantivalist approaches within Greek philosophy. We consider the relational theory of Plato as presented in {\it Sophist} and the substantivalist theory of atoms as presented by Democritus and Leucippus. Section 3 reconsiders the  relational-substantivalist debate in Modernity, more in particular, the relational scheme proposed by Spinoza and the triumph of the atomist metaphysics through its implementation in Newtonian physics. In section 4 we will introduce quantum theory as deeply related, in the 20th century, to both atomism and positivism. In section 5 we will discuss several relational accounts of quantum mechanics such as those of Bohr, Bene-Dieks and Rovelli. Finally, in section 6, taking into account our previous analysis, we will present an ontological account of relations with which we attempt to provide a new approach for representing the theory of quanta. 

\section{Philosophy, Physics and Sophistry: Realism or Relativism?}

Let us remember once again the Greek moment, the origin of both physics and philosophy. And let's remember, to emphasize this common origin, the name that Aristotle uses to refer to the first philosophers: the ``physicists''. This denomination comes from the object that, according to multiple sources, they all intended to describe: {\it phúsis}. A term that is unanimously translated as ``nature'' and whose meaning covers what we refer to when we talk about ``the nature of reality'' (its essence), as well as what we commonly, broadly and in an extensive way refer to as nature: the reality in which we take part. {\it Phúsis} is, for physicists, something dynamic which ---at the same time--- responds to some sort of internal order or formula. Some of these first philosophers proposed an ``element'' (or a series of them) from which ---and according to which--- all reality develops and can be explained. But among them there are also others who didn't follow this strategy. In particular, there are two of them who are particularly important for our analysis: Heraclitus of Ephesus and Parmenides of Elea.  

Heraclitus redirected the search for the fundament of {\it phúsis} no longer to an ``element'' but to the description of a formula, an internal order that rules {\it phúsis}. He described this formula and called it {\it lógos}. This denomination is very significant for the development of philosophy. Until Heraclitus' use of the word, logos had a meaning exclusively related to language: discourse, argumentation, account, even tale. In all of those translations we can see already something that will be essential to all meanings and nuances of logos, even when it doesn't refer to language: a significant combination, a reunion with criterion, a collection with purpose. {\it Lógos} never means an isolated word, or a meaningless sentence, or dispersed and ineffective ensembles of words. It always refers to a combination that is able to produce an effect or a meaning. We now begin to understand why Heraclitus chooses this specific word to name the internal order of phúsis. He sees in {\it phúsis} exactly that: a combination that responds to a formula, a criterion. This double meaning of {\it lógos} ---formula of {\it phúsis} and human discourse--- implies an affinity between language and reality that allows for philosophical knowledge: it is in a linguistic manner that we are capable of exposing the internal order of reality. Thus, there is an affinity between the {\it lógos} of men and the lógos of {\it phúsis}. However, it is a difficult task to expose the true {\it lógos} since, as remarked by Heraclitus, ``{\it phúsis} loves to hide.'' [f. 123 DK]. Doing so requires hard work and sensibility, but ---following Heraclitus--- the latter can be revealed in the former. In a particular {\it lógos} one can ``listen'' something that exceeds it, that is not only that personal discourse but the lógos of {\it phúsis}: ``Listening not to me but to the {\it lógos} it is wise to agree that all things are one'' [f. 50 DK]. We are thus able to represent {\it phúsis}, to exhibit its {\it lógos}.

In another part of the ancient Greek world, Parmenides makes a discovery that, as all great discoveries in the history of philosophy, is beautifully simple: {\it there is being (and not being is impossible)}. No matter the ``element'' or the formula that you may choose as fundamental for exposing {\it phúsis}, the truth is that anything, any element, order, etc., must necessarily and ``previously'' {\it be}. The ``fact of being'' ---according to Néstor Cordero's formula \cite{Cordero05}--- comes before any determination that we can predicate of whatever. Parmenides' philosophy begins with the evidence of this all encompassing and irreducible fact of being, and searches what we can say starting from it. As Heraclitus, and maybe in a more explicit manner, he will affirm a natural relation between being, thought and language; a relation that will become, at least for some years, a sort of dogma.

But time passes, things change, and philosophy, already in its youth, will encounter its first opponent, its first battle. We are in the V century before Christ and Athens is the most powerful (politically, culturally and intellectually) {\it polis} of the Greek world. Thinkers from all over the region come to Athens. Some of these foreigners begin to make a living by teaching Athenian citizens. But, unlike physicists and philosophers, they don't teach how to know the true nature of reality; instead they teach techniques of argumentation and persuasion, techniques of great utility in the {\it \'agora}. In private lessons and public conferences, the sophists prove the persuasive power of {\it lógos}, understood now exclusively as human discourse ---and independent of the {\it lógos of phúsis}. It is quite a blow for philosophy, which in a somewhat naïve way had affirmed a privileged, unbreakable relation between truth and {\it lógos}. But, contrary to what their fame indicates, it is not only the taste for controversy what lies behind sophistry, there are some originally sophistic positions ---which evidently arose from an opposition to philosophy as it was known--- that justify their {\it praxis}. Some strongly sceptical postures. These are positions that undermined the basis of the philosophical attempt to represent {\it phúsis}. In order to develop the fundamental aspects of these positions, it is useful to focus on two of the most famous and prodigious sophists: Protagoras and Gorgias. From the former an original phrase remains ---that will be used many centuries later by the Vienna Circle in their manifesto--- which has become famous: ``Man is the measure of all things, of the things that are, that they are, of the things that are not, that they are not'' [DK 80B1] . We don't possess much more of Protagoras' text but we do have some comments about his philosophy that date back to antiquity, and they all seem to coincide: he proposes a relativistic view. According to this stance, there is no such thing as `a reality of things' ---or at least, we are not able to grasp it. We can only refer to our own perception. Things do not have a reality independent of subjects; and even in the case such a reality would exist we simply cannot have access to it. We, individuals, have only a relative knowledge dependent of our perception: ``Italy is beautiful ---for me''. This is a stance that has undoubtedly an intuitive appeal and that we will encounter again in some of the ---more contemporary--- positions we will analyze in this article. Gorgias, on the other hand, a quite talented orator, does a systematic critique of the founding principles of philosophy ---in particular of the Parmenidean philosophy. In his famous discourse, {\it On Not Being}, he tries to dismantle the relations that Parmenides establishes between being, thought and {\it lógos}. A brief summary of the discourse would be: there is not a being of things (or even: ``nothing exists''); even if there is a being of things, it would not be accessible to our thought; and finally if our thought would be able to grasp the being of things, we still wouldn't be able to communicate it. 

Evidently, philosophy could not continue without resolving, or at least responding, the deep arguments of the sophists. The attempt to save and precise the nature and method of philosophical knowledge will give birth, among other things, to two of the most important works of western thought. We refer ---of course--- to the philosophies of Plato and Aristotle. Even if the positions of sophistry and philosophy are in great measure irreconcilable, and if this opposition then continues ---more or less implicitly, as we will see--- throughout the history of thought, the truth is that sophistry has accomplished an important role in the history of philosophy for it constitutes the first critical moment that has proven to be fundamental for the development of philosophy itself.

\section{Relationalism vs. Substantivalism in Greek Thought}

We will come back to the reappearance, within the history of Western thought, of the strong critical arguments that gave sophistry its importance, but first we want to establish another opposition that appears inside philosophy (and physics), this is, an opposition between two ways of representing {\it phúsis}, two fundamentally different views regarding reality. Both confident in the human capability of expressing {\it phúsis} through a discourse, but each one arriving at a very different representation of the world. We are talking about substantivalism and relationalism. The former is a view of the world as populated by multiple individual substances, the latter refuses to introduce ontological separation within reality and defines its elements as relations ---not as separated existents. We want to clearly establish which are the fundamental differences between these two worldviews so that we can produce a solid basis for the discussion and analysis of what should be considered the substantivalism-relationalism debate. 

\subsection{Plato's Relationalism in the Sophist}

The interest in relationalism is relatively recent. During the last century, different developments in various disciplines seem to tend towards the need of a relational understanding of reality. But the truth is that, if we look back, we find that in the history of both physics and philosophy there have always been some doctrines that we might catalog as ``relationalist'' ---even if they are not explicitly described in that manner. We can even say that what we will call here ``substantivalism'' ---the view opposed to relationalism---, that is, the conception according to which the world is made of individual and independent separated substances, has been seen in philosophy with suspicious eyes. The foundation of this view was recognized as weak or even unsustainable, even if some of its characteristics might seem to coincide today with a ``common sense'' understanding of reality.

From a chronological point of view, the first philosopher who defined being as relation (more specifically: as capacity or power of relation) was Plato. Returned from his second trip to Sicily, already an old man but more lucid than ever, Plato transforms his own philosophy. This is especially visible in a series of three dialogues which conclude with the difficult yet prodigious {\it Sophist}. First, in the {\it Parmenides}, Plato uses the figure of the Eleatic philosopher to carry out a critique of an orthodox version of the `Theory of Ideas' ---the one we can find in his previous dialogues. Through a fictitious Parmenides, Plato does some autocriticism, leaving the character of Socrates ---in a very young version--- mourning the loss of his beloved, and now refuted, Theory of Ideas. But Parmenides tries to fight Socrates' anguish with some words of encouragement: he praises the Socratic attempt to direct his search for knowledge towards the intelligible and he gives him confirmation that, despite his previous failure, it is necessary ---if we do not want to discard knowledge as impossible--- to sustain the existence of Ideas. This means we still need a Theory of Ideas, just not the same of the previous dialogues. In the next dialogue, the {\it Theaetetus}, Socrates searches the possibility of defining true scientific knowledge ({\it epistéme}) without postulating Ideas. All attempts fail, and the characters of the dialogue decide to meet again next morning for the continuation of the discussion. What happens the next morning is what the {\it Sophist} describes. Socrates, Theaetetus and Theodore meet up, and one of them has invited a new participant: the mysterious Eleatic Stranger. This Stranger will be Plato's spokesman for the majority of the dialogue, and he will expose the new version of the Platonic Theory of Ideas. Meanwhile, Socrates listens in silence. It is in this context that Plato will define being, for the first time in his philosophy. He will define it as {\it dunamis} of affecting and being affected, {\it dunamis} of communicating, {\it dunamis} of relation. Being means to posses this {\it dunamis} (see \cite{Fronterotta95, Gonzalez11}). This Greek term is usually translated as potency (especially when it is Aristotle who uses it), possibility, capacity, power. In each one of these translations we emphasize either the passive or the active aspect which coexist within the Greek term. But before exploring the meaning of this definition, let us begin by pointing out something evident: the transformation of the platonic philosophy that we witness in these dialogues has a strong Eleatic background. Let us remember that it is Parmenides who, speaking in the name of Plato, refutes the previous Theory of Ideas, and afterwards it is also an Eleatic who accomplishes the task of changing the theory. Undoubtedly Parmenides is, next to Socrates, the philosopher that most influenced Plato. The pejorative characterizations, more or less ironic, that Plato uses to describe the thought of the previous and contemporary philosophers, are left asides when he talks about Parmenides. He shows him respect, sometimes veneration (even if he is forced, in the {\it Sophist}, to contradict him in some aspects). It is useful then to briefly remember Parmenides' philosophy. His main thesis can be simply expressed (but not so simply interpreted): {\it there is being (or `there is what is', or `there is the fact of being') and not being is impossible.} If the previous philosophers, or those contemporary to Parmenides, gave privilege to one or several `elements' as origin and foundation of nature, or dedicated their thought to decipher the hidden order that governs reality (the case of Heraclitus), Parmenides starts by reflecting on a previous truth: any `element', any `order', anything of any nature, must be, first, something that is. The simple yet universal, all encompassing fact of being is the origin of the Parmenidean wonder. And as evident as the fact of being is the impossibility of its contrary: not-being. One of the ways in which Parmenides phrases this impossibility is the one that identifies not-being with separation: ``you will not sever what is from holding to what is'' [f. 4]; ``it is wholly continuous; for what is, is in contact with what is'' [f. 8.25]; ``Nor is it divisible, since it is all alike'' [f. 8.22]. In this sense, ``There is being and not-being is impossible'' means that there is no cut, no strip, no ditch, inside being ---through which not-being would pass. Being has no cracks within, no interstices. But how should we then continue? What else to say besides ``there is being''? Which content can we predicate of the fact of being? It seems ---at first sight--- like an empty discovery. Plato, with a truly parmenidean spirit (despite the parricidal declaration in the {\it Sophist}), tries to go a little further, or to be a bit more specific: he proposes a definition of being as {\it dunamis}. ``And I hold that the definition of being is simply {\it dunamis}'' [{\it Sophist}, 247e]. Plato says that being is the {\it dunamis} to act and be acted upon, the capacity to affect and be affected. The fundamental reality of everything is, for Plato, this inherent tendency to relation, this potency of relation. Néstor Cordero, in a commentary to the {\it Sophist} \cite[p. 155]{Cordero14}, describes this {\it dunamis} as ``the capacity of an entity, any entity, for relating with another (either affecting or being affected) and for that reason a few pages later Plato replaces `acting' and `suffering' by a single verb, `communicating', and he talks about {\it dunamis koinonas}, `the possibility or potentiality of communication' [{\it Op. cit.}, 251e]. And since being is communicating, something that doesn't communicate doesn't exist. (...) Plato assimilates this potentiality to the fact of being and he gives precisions about it: it is the possibility of communicating, that is, (...) to produce reciprocal bonds''. This being in everything that is, it is {\it dunamis} of affecting and being affected, of interacting, it is an inherent tendency towards relation. It is the participation of a universal relationability. If all things, from any kind, are, and ---according to Parmenides' lesson--- there can't be not-being between them, if always ``what is, is in contact with what is'', it is then ---Plato adds--- this same universal communicability, this unbreakable basic tendency towards relation, the nature on the fact of being. Nothing can exist, for Plato, which does not possess this {\it dunamis}. Anything, any ``something'' is, first, potency of relation, impossibility of being something isolated, it is part of a universal communicability. Where there is potency of relation there is being, and vice versa. For the first time we encounter a relational ontology. Parmenides, with good reasons, denied the possibility of identifying being with a given qualification, because this would relativize or limit the fact of being. But Plato finds a qualification that can be applied to being without limiting or relativizing it, that leaves nothing outside being, that doesn't identify being with a determined kind of entity, that every existent, of every kind, shares: the capacity of relation. This capacity, like Parmenides' being, has no possible contrary. There are no possible ontological separations. It is then a universal qualification, with no opposite, without limits, and which also allows for a more specific knowledge of reality, of which is given among being. It restores its dunamism and variety without introducing not-being, without postulating separated substances. Because not every relation is the same relation; relations have variable intensities, different qualities; sometimes there is affecting, sometimes there is suffering. Relation allows us to articulate the general ---and immutable--- truth of being (and the absence of not-being) with the specific ---and mutable--- reality of experience. 

Although this deduction from the Parmenidean fact of being to Plato's being as {\it dunamis koinonas} ---following the guide of the impossibility of separation--- is reasonable when following the mentioned texts, the truth is that Plato does not describe his path towards the definition of being in this manner. Instead, he does so by putting together a brief revision of the previous philosophies; what might be considered to be the first ``history of philosophy'' within the history of philosophy. By this revision he tries to discover what ``being'' could mean. He wonders: which characterizations can be extracted from previous philosophies? What he finds is that the past philosophers mistake being with some particular determinations, either quantitative or qualitative ones, that limited it. Determinations that cannot bear the universal applicability (and absence of contrary) that corresponds to being. Some identified what is to a determined opposition, a duality, as for example the hot and the cold. Others identified being with the One (denying multiplicity). Then he verifies that some of them referred being to material things, and others (Plato himself, before the {\it Sophist}) identified it with intelligible entities. All of them qualified or quantified being, they identified being as some kind of limited determination. And that created problems for them. They always seemed to leave some things outside being. What Plato is proposing to them is that they should expand, broaden, their conception of being, that they should not limit being to a particular kind of entity, excluding from existence ---for that reason--- things that also exist. In the context of the discussion with the materialist (who denied incorporeal realities) he proposes a solution, a settlement: let's just say that all that has the possibility of interacting exists. ``Anything which possesses any sort of {\it dunamis} to affect another, or to be affected by another, if only for a single moment, however trifling the cause and however slight the effect, has real existence; and I hold that the definition of being is simply {\it dunamis}'' [{\it Op. cit.}, 247e]. The lesson that we can extract from this brief history of philosophy is the necessity of not limiting being to a specific type of entity, or to certain number of entities. Instead, we must conceive being in a way inclusive enough so that everything that proves itself existent, is covered by such proposed metaphysical account. And what we undoubtedly know about everything that is, is that, in a greater or smaller degree, it is communicated with the rest, it has relations, it is included in a universal communicability. Relation is thus irreducible and primary. 

However, there is still a point in which the Parmenidean heritage remains uncomfortable. We need to be able to say that everything is, that the fact of being is universally applicable, yes, but at the same time we also know that everything is not the same. Experience tells us that being is full with differences. One of the main problems for a truly relational conception of being, one that denies ontological separation and takes relation as fundamental, is to justify the differences inside being without producing separated substantial individuals. Here enters the main concept of the Sophist: {\it difference.} There is not an absolute not-being, the contrary of being does not exist, there are no separations, but there are differences. These differences don't amount to separations (in fact difference is a relation), but they do account for variety and multiplicity among being. Plato introduces in this manner ---and for the first time in the history of philosophy--- a relative notion of not-being. There is no absolute not-being, no void, the nothing does not exist, but everything that is, at the same time is not an infinity of other things (it is different from them). This table, for instance is not (is different from) a chair. Difference, as well as Sameness, and some other determinations, articulates the identity and specificity of things that exist, but this only on the irreducible base of being, that is to say, on the base of the potency of relation that subtends, comprehend and renders them dynamic.

\subsection{Greek Atomism: a Substantivalism}

Among the first philosophers a particular school formed to which we now turn our attention. It began with Leucippus (of whom we have no original texts) and his pupil Democritus (of whom we have several original fragments from his apparently numerous books). They proposed, as many of the first philosophers, some fundamental elements out of which {\it phúsis} was made of. These elements where being and not-being, which in term they interpreted as `the full' and `the void'. Contrary to Parmenides, they conceived the existence of a non-being which is the contrary of being. In their minds, there is void within nature. Being, or the full, consisted to them of indivisible bodies, indivisible fragments of mass with a minimum size. Simple bodies. They used an adjective to describe these bodies: {\it átomos}, which means, literally, ``not divided''. That adjective became an -ism and this school was called ``atomism''. 

Atoms couldn't be infinitely small, there's a limit to how small they are. If not, atomism wouldn't work, we wouldn't be able to say that ``atoms have mass'', and we would be stuck with a difficult conclusion for them (a conclusion that Zeno of Elea already pointed out astutely): bodies would be made of zeros, mass would be made of not-mass. A conclusion that for atomism would be catastrophic. The admission that atoms are indivisible bodies with a small amount of mass is for atomists an axiom. They don't justify it, they only postulate it. They sacrificed the justification of this admission in order to produce an explanation of reality that seemed coherent to them. It is easy to demonstrate that this axiom is problematic, for it entails the existence of a mass that can't be divided. If we want to be fair, we can say that Parmenides also started from a postulate he wouldn't justify: there's being and not-being is impossible. If we compare the amount of presuppositions made by one and the others: Parmenides presupposes being (and the inexistence of not-being), while the atomists presuppose being, not being, and that being is made of indivisible small bodies with mass that travel in void. But let's leave the comparison aside, for it is not the economy of presuppositions that interests us. For atomists, then, everything is made of these simple bodies that move around in void. Atoms have shapes, and according to these shapes they can unite to form more complex bodies. In conclusion, what it's important to us is that atomism is without a doubt a substantivalism: there are small individual substances, and these substances are separated from each other by not-being. The world for atomists is made of these separated substances.

\section{Relationalism vs. Substantivalism in Modernity}

The tension between relationalism and substantivalism remained through the history of western thought, sometimes in more central arenas, sometimes in more marginal ones. However, there is in modernity a triumph of substantivalism which would change the balance between these two opposite accounts of reality. Even though relationalism still remained an important viewpoint within modernity, the power of Newtonian mechanics interpreted on the lines of atomism would determine the fate of what would be later on recognized as ``classical common sense''. But before considering the triumph of the newborn ``mechanical atomism'' imposed by Newton, let us begin by recalling what we consider to be a particular interesting development on the lines of relationalism. A development which will prove particularly interesting for the discussion regarding the meaning and understanding of relations.

\subsection{Spinoza's Relationalism}

If we would pay attention only to terminology, it wouldn't seem logical to pick Spinoza's philosophy as our next example of relationalism, since he intends explicitly an analysis of substance. But one always has to pay attention to what each philosopher does with the terms he uses within his own philosophy, what does he takes from the traditional meaning, what does he change. This is particularly useful when faced with Spinoza, since it is evident ---even for those who, in his own time, didn't understand him--- that he took the terminology of Cartesian and scholastic traditions, but in order to say ---with those same terms--- something completely different. The most emblematic case is, undoubtedly, the term ``substance''. The meaning is unprecedented: for Spinoza there are no multiple substances, there is only one single substance, with infinite attributes. And there's nothing else besides it. It is, in general terms, an equivalent of the Parmenidean being. All the variety that we encounter is the infinite variety of modifications of only one substance, according to its different attributes. But before analyzing the nature of these modes of the substance (where relation appears), let us see, even in a brief manner, how unity and continuity of substance ---and impossibility of separation--- represent the basic aspects of his ontology. 

Spinoza's account of nature pictures a similar landscape to the one we drew from Parmenides and Plato, but now developed with a modern terminology: attribute, mode, real distinction, modal distinction, etc. For Spinoza, nor the difference between attributes (extension, thought), nor the difference between modes (things, bodies, ideas, souls, etc.), entail substantial divisions. There aren't any differences, in modes or in attributes, which imply separations. For there to be separations there would have to be distinct substances, and Spinoza proves, especially in the first book of the {\it Ethics}, that ---first--- there are no multiple substances of the same attribute, and ---second--- that there is only one substance for all attributes. For Spinoza there's no separation, no not-being, no void in reality. It is true that the differences between attributes are real differences, but these don't amount to substantial distinctions, only to qualitative differences. Attributes are the qualitative natures according to which the same substance expresses itself. Also modal differences don't amount to separations, only to the different modifications inside the one and only substance: ``As regards the parts in Nature, we maintain that division, as has also been said before, never takes place in substance, but always and only in the mode of substance'' [{\it Short Treatise I}, chap. II, 19-22].

But physics is not the science of {\it being qua being}, nor of God as an absolutely infinite substance, but ---to put it in Spinozian terms--- the science of the modes of the substance. As we have seen, modes are not substances but modifications of a unique substance, according to its different attributes. One of the main issues of a non-substantivalist ontology is ---as we said before--- to give an account of the singularity of each individual (the many) without producing separated substances (the one). This means arriving to the multiple individuals without losing the unity of being. What defines an individual, that which essentially characterizes its singularity and distinguishes one from the others, is ---according to Spinoza--- a relation. A ``part'' of the divine potency, or in other words, a degree of potency that expresses itself as a relation. Thus, following the physics of his time, each individual is characterized as a specific relation of movement and rest. Spinoza tells us that these relations can be larger or smaller, more or less ``perfect''. There are differences among them, but these differences can't be thought of as an extensive quantity (e.g. the quantity of mass). There is still a quantitative difference that distinguishes relations, but it is no longer an extensive quantity, it is an intensive one, a degree of potency. In the order of relations that don't depend on their terms ---in the context of a relational ontology--- what distinguishes and quantifies different relations are their intensities of potency. 
Individuals, in Spinoza's philosophy, are essentially potency and relation. They are quantities of potency because they are modifications, of the same univocal being. They are relations because those potencies are expressed in relations, different relations, specific relations. And, according to Spinoza, from simpler to more complex relations, made of the composition of those simpler relations, we arrive at more complex individuals. And we also arrive to the idea that the totality of the relations between these relations would give us the totality of nature. An individual composed of all individuals. The total relation of relations. Since separation is not possible, within a relational ontology we do not have, we can't have, individual substances. What we have instead are modifications of being, elements that can be different from each other but that can't truly be separated from each other. But still, every one of those individuals can be described, can be quantified, can be experienced as a particular individual, we are not lost in an indefinite totality.

\subsection{Newtonian Physics and the Triumph of Atomism}

It was Isaac Newton who was able to translate into a closed mathematical formalism both the ontological presuppositions present in Aristotelian logic together with the materialistic reduction of reality to {\it res extensa} ---taking in this way actuality as the unique mode of existence of things. He did so with the aid of atomistic metaphysics. In the V and IV centuries B.C., Leucippus and Democritus had imagined existence as consisting of small simple bodies with mass. According to their metaphysical theory, atoms were conceived as small individual substances, indivisible and separated by void. The building blocks of our material world. Many centuries later, Newton had been able not only to mathematize atoms as points in phase space, he had also constructed an equation of motion which allowed to determine the evolution of such ``elementary particles''. The picture of the world described by Newtonian mechanics was that of small completely determined particles bouncing between each other in space in an absolutely deterministic manner. The obvious and most frightening conclusion implied by the conjunction of Greek atomism and Newton's use of the effective cause was derived by the mathematician Pierre Simon Laplace:

\begin{quotation}
\noindent {\small ``We may regard the present state of the universe as the effect of its past and the cause of its future. An intellect which at a certain moment would know all forces that set nature in motion, and all positions of all items of which nature is composed, if this intellect were also vast enough to submit these data to analysis, it would embrace in a single formula the movements of the greatest bodies of the universe and those of the tiniest atom; for such an intellect nothing would be uncertain and the future just like the past would be present before its eyes.'' \cite[p. 4]{Laplace}}
\end{quotation}

In the XVII Century, in the newly proposed mechanical description of the world, the very possibility of the indetermination supposed by the potential realm of being had been erased from physical reality. In classical mechanics, every physical system may be described exclusively by means of its actual, coexistent (in a non-contradictory way) and determined properties. A point in phase space is related to the set of values of properties that characterize the system. In fact, an actual property can be made to correspond to the set of states (points in phase space) for which this property is actual. Thus, the change of the system may be described by the change of its actual ---meaning, preexistent or independent of observation--- properties. Potential or possible properties are then considered as the points to which the system might (or might not) arrive in a future instant of time. Such properties are thought in terms of irrational potentiality; as properties which might possibly become actual in the future. As also noted by Dieks: 

\begin{quotation}
\noindent {\small ``In classical physics the most fundamental description of a physical system (a point in phase space) reflects only the actual, and nothing that is merely possible. It is true that sometimes states involving probabilities occur in classical physics: think of the probability distributions in statistical mechanics. But the occurrence of possibilities in such cases merely reflects our ignorance about what is actual. The statistical states do not correspond to features of the actual system (unlike the case of the quantum mechanical superpositions), but quantify our lack of knowledge of those actual features.'' \cite[p. 124]{Dieks10}}
\end{quotation}

Classical mechanics tells us via the equation of motion how the state of the system moves in phase space along the curve determined by the initial conditions and thus, any mechanical property may be expressed in terms of phase space variables. Needless to say, in the classical realm the measurement process plays no role within the description of the state of affairs and actual properties fit the definition of elements of physical reality in the sense of the EPR paper [41]. Moreover, the structure in which actual properties may be organized is the (Boolean) algebra of classical logic. With Newtonian physics, modernity embraced ---at least for the physical realm--- a substantivalist and materialistic representation of the world commanded by the efficient cause. A view that was nourished by atomism, by Aristotle's both logical and ontological principles of existence, non-contradiction and identity, and by the reduction of existence only in the restrictive terms of the actual mode of being. A physics of pure actuality. The Newtonian metaphysical representation of the world as an ``actual state of affairs'' remained a dictum that still traverses not only classical physics, but also relativity theory. It was only the appearance of the theory of quanta that disrupted the classical ---actualist and atomist--- representation of the world, producing a revolution that ---as Constantin Piron \cite{Piron99} has remarked--- has not yet fully taken place. If classical physics has sustained for quite some time the limitation of what exists to actual substantial entities, quantum physics came to break that limitation, and forces us now ---following the example of the Eleatic Stranger--- to consider the broadening of our understanding of reality maybe even beyond substantivalism and the actual mode of existence. But to apprehend the development of the physics of quanta we must understand how Machian positivism, by way of a deconstruction of the Newtonian {\it a priori} notions of absolute space and time, was able to place physics within a new critical moment.

\section{Quantum Theory: Between Atomism and Positivism}

Positivism was born in the XIX century, taking elements from both English empiricism and French Enlightenment. On the one hand, in a reaction against metaphysics, it stood on the idea of founding knowledge on sensible data; on the other hand, it maintained a generalized trust on the progress of reason and science. Positivism derived from thinkers like Laplace and many others, but was first systematically theorized by August Compte, who saw in the ``scientific method'' the possibility of replacing metaphysics in the history of thought. Just a century before, Kant had also fought what he saw as the ``dogmatic'' metaphysics of his time, developing a new system capable not only of resolving the dispute between rationalists and empiricists of the XVII and XVIII centuries, but also capable of justifying Newtonian physics as ``objective knowledge''. However, a century later, the categories and forms of intuition had become ---according to many--- exactly what Kant had striven to attack: dogmatic and unquestioned ideal elements of thought. Against metaphysics, positivism stated that the only authentic knowledge is knowledge that is based on actual sense experience. Such knowledge can only come from the affirmation of theories conceived in terms of what was believed to be a ``strict scientific method''. Metaphysical speculation ---understood now as a discourse attempting to go beyond the observed phenomena--- should be always avoided and even erased from scientific inquiry and research. 

Ernst Mach is maybe one of the most influential positivist thinkers of the XIX century. His criticisms might be regarded as the conditions of possibility for the development of physics that took place at the beginning of the XX century. He developed a meticulous deconstruction of the fundamental concepts of Newtonian physics; a critique that produced a crisis in the fundament of scientific thought itself. This crisis was certainly a standpoint not only for the birth of relativity ---as recognized by Einstein himself--- but also played an essential role in the development of the theory of quanta. Mach, a physicist himself, was primarily interested in the nature of physical knowledge. His investigations led him to the conclusion that science is nothing but the systematic and synoptical recording of data of experience. In his {\it Analysis of Sensations}, Mach concluded that primary sensations constitute the ultimate building blocks of science, inferring at the same time that scientific concepts are only admissible if they can be defined in terms of sensations.
 
\begin{quotation}
\noindent {\small ``Nature consists of the elements given by the senses. Primitive man first takes out of them certain complexes of these elements that present themselves with a certain stability and are most important to him. The first and oldest words are names for `things'. [...] The sensations are no `symbols of things'. On the contrary the `thing' is a mental symbol for a sensation-complex of relative stability. Not the things, the bodies, but colors, sounds, pressures, times (what we usually call sensations) are the true elements of the world.'' \cite{Mach}}
\end{quotation}
 
\noindent In Machian positivism there is thus no room for {\it a priori} concepts, nor for unobservable entities ---like atoms. Talking about entities that can't be observed is to fall in the trap of metaphysics, to go beyond phenomena producing a discourse with no meaning nor reference, to detach our discourse from the only possible true reference: sensations. In what was one of the main scientific controversies of his time, Mach firmly opposed to accept the existence of atoms. However, as a result of the experimental and theoretical work developed by physicists like Dalton, Maxwell and Boltzman, towards the end of the XIX century, atomism ---not without resentment of a newborn community which went back to a wave type description of reality--- had won the battle and occupied once again the dominant position in scientific communities. It is in this same context that the theory of quanta would make its appearance, producing very soon a paradoxical entanglement between two mutually incompatible positions, namely, atomist substantivalism ---that maintained, in metaphysical terms, the existence of unobservable atoms--- and Machian positivism ---which grounding itself in observed phenomena affirmed the need to eradicate all metaphysical notions from physics, including of course that of ``atom''.

Quantum physics was born together with the XX century, after the introduction by Max Planck in 1900 of the ``quantum postulate'' ---in order to solve a problem related to the emission of radiation by hot bodies. Quantum theory begun its history as a theory about atoms. Its development continued through the first three decades of the XX Century, when it finally became what we know today as ``Quantum Mechanics''. But once the formalism of quantum mechanics had become a closed mathematical scheme, it also became very soon evident to the founding fathers of the theory that there were too many problems to conceive the theory as describing physical reality in terms of atoms ---as ``tiny elementary particles living in space-time''. According to Heisenberg \cite[p. 3]{Heis58}, ``the change in the concept of reality manifesting itself in quantum theory is not simply a continuation of the past; it seems to be a real break in the structure of modern science''. Quantum contextuality, the existence of strange superpositions, the measurement problem, the problem of quantum individuality and the problem of non-locality, among many others, showed the limits of attempting to understand quantum physics in terms of an atomist ontology. 

Concomitant to quantum mechanics, in the first decades of the XX century logical positivism was also developed attempting to fight metaphysical thought through the development of Mach's ideas and his empiricist standpoint. Congregated in what was called the Vienna Circle, in their famous manifesto \cite{VC} they argued that: ``Everything is accessible to man; and man is the measure of all things. Here is an affinity with the Sophists, not with the Platonists; with the Epicureans, not with the Pythagoreans; with all those who stand for earthly being and the here and now.'' Their main attack to metaphysics was based in the idea that one should focus in ``statements as they are made by empirical science; their meaning can be determined by logical analysis or, more precisely, through reduction to the simplest statements about the empirically given.'' Their architectonic stood on the distinction between ``empirical terms'', the empirically ``given'' in physical theories, and ``theoretical terms'', their translation into simple statements. This separation and correspondence between theoretical statements and empirical observation left aside metaphysical considerations, regarded now merely as a discourse about un-observable entities, pure blabla. One of the major consequences of this empiricist perspective towards observation is that physical concepts become only supplementary elements in the analysis of physical theories. At most, an economy to account for physical phenomena. When a physical phenomenon is understood as a self-evident given (independent of physical concepts and metaphysical presuppositions), empirical terms configure an objective set of data which can be directly related ---without any metaphysical constraint--- to a formal scheme. Actual empirical observations become then the very fundament of physical theories which, following Mach, should be understood as providing an ``economical'' account of such observational data. As a consequence, metaphysics, understood as a conceptual and systematic representation of {\it phúsis}, was completely excluded of the main positivist picture attempting to describe scientific theories.

\begin{center} 
{\it Empirical Observable Data --------------- Theoretical Terms

\bigskip 

(Supplementary Interpretation)}
\end{center}

According to this scheme, physical concepts are not essentially needed, since the analysis of a theory can be done by addressing only the logical structure which accounts for the empirical data. The role of concepts becomes then accessory: ``adding'' metaphysics might help us to picture what is going on according to a theory. It might be interesting to know what the world is like according to an interpretation of a formalism but, as remarked by van Fraassen  \cite[p. 242]{VF91}: ``However we may answer these questions, believing in the theory being true or false is something of a different level.'' The important point we would like to remark here is that according to this empiricist viewpoint, since the world is unproblematically ``described'' in terms of our ``common sense'' understanding of phenomena an adequate empirical theory can perfectly account for experiments without the need of an interpretation.\footnote{It is important to remark that the problem of interpretation in the context of philosophy of physics has been a deep problem since its origin. The relation between empirical observation has been a difficult subject of analysis since Carnap, Neurath, Popper, Hempel and many others tried to escape the metaphysical characterization of physical concepts. See \cite{Cassini17}.} However, the project of articulating the empirical-formal relation through the distinction between {\it theoretical terms} and {\it observational terms} never accomplished the promise of justifying the independence of those realms ---specially with respect to categorical or metaphysical definition of concepts. The fundamental reason had already been discussed by Kant in the {\it Critique of Pure Reason}: the ``actual observations'' (or empirical terms) can't be considered as ``givens'', the observation cannot be understood nor considered without previously taking into account a categorical structure that allows to account for phenomena. The description of phenomena always presupposes  ---implicitly or explicitly--- metaphysical elements. Identity or non-contradiction are not ``things'' we see in the world but rather the very conditions of possibility of classical experience; we presuppose them in order to make sense of the world. This categorical systematization, allowing for a theoretical-conceptual representation, is in itself metaphysical. As the philosopher from K\"onigsberg would have said, it is the representational framework of the transcendental subject, articulating categories and forms of intuition, that which allows for an objective empirical experience. 

After the second World War, and establishing a continuity with positivism, the Anglo-Saxon thought consolidated in what was called ``analytical philosophy''. This new philosophy was originated in opposition to another supposed philosophical ``school'', called ``continental'' ---meaning the European continent. Even if the branching of analytical philosophy advanced in a vertiginous manner in the academic world, and even with the internal critiques in the 60's and 70's by figures as Lakatos, Feyerabend and Kuhn (among others), we can safely say that the fundamental presuppositions remained those of classical logical positivism. Contemporary philosophy of science (not only of physics) continues to rely on two fundamental distinctions: one between ``theoretical terms'' and ``empirical terms'', and the other between ``observables'' and ``unobservables''. About the first of those distinctions, Curd and Cover \cite[1228]{PS} affirm: ``Logical positivism is dead and logical empiricism is no longer an avowed school of philosophical thought. But despite our historical and philosophical distance from logical positivism and empiricism, their influence can be felt. An important part of their legacy is observational-theoretical distinction itself, which continues to play a central role in debates about scientific realism.'' And about the distinction between observables and unobservables, Musgrave \cite[1221]{PS} explains: ``In traditional discussions of scientific realism, common sense realism regarding tables and chairs (or the moon) is accepted as unproblematic by both sides. Attention is focused on the difficulties of scientific realism regarding `unobservables' like electrons.'' This perspective has very deep consequences for research not only in philosophy but also in physics. In particular, it closes the door to the development of radically new physical representations, since it assumes that we already know what reality is in terms of the (naive) ``common sense'' observation of tables and chairs ---also known, following Sellars, as the ``manifest image of the world''. It is in this frame that the problem of realism has been reconfigured ---inside the limits established by the newborn philosophy of science--- around the question of the scientific justification of a ``given'' reality, exhibited always through the ``common sense'' language that we use to give an account of what we observe, and not around the question of the means to produce a systematic theoretical representation or expression of reality. Realism is then situated inside the limits imposed by a perspective according to which it is the ``self-evidence'' of what is observed by individuals ---and not the representation of {\it phúsis}, of reality--- the true fundament of knowledge.

It is by these multiple paths that we arrive to the current situation in which the philosophy of quantum mechanics is at the center of a perfect storm created around the questions of its meaning and reference. These questions are articulated in a paradoxical manner, by sustaining two mutually incompatible perspectives, in what we could call a curious ``sophistic substantivalism''. On one hand, the philosophy of physics tries to produce a bridge between, first, a language assumed by physicists in terms of unobservable elementary particles ---namely, the atoms, protons, electrons, quarks, etc.---, and, secondly, a ``common sense'' language where ``tables and chairs'' are taken a-critically as ``self-evident'' unproblematic existents. On the other hand, the referentiality of theories is considered under a double standard, where we still consider science as an economy of the experience of subjects (of experiments and measurement outcomes), and at the same time we ask ---with little conceptual support--- about the reality of the world beyond measurement results. So it seems, the very foundations of the project rest on the paradoxical entanglement of a substantivalist metaphysics that refers to unobservable particles, with an observational empiricism that, while aiming at leaving aside the metaphysical-conceptual debate, tries to justify at the same time the existence of our ``common sense'' (but still metaphysical) classical representation of the world.

\section{Epistemic Relationalism in Quantum Mechanics}

What might be called in a broad sense ``the epistemic view of quantum mechanics'' has become one of the main viewpoints accepted not only within philosophy of quantum mechanics, but also within physics itself. According to this perspective, in line with empiricism, observation is not considered as something problematic. Observation is considered not only as the ground but also as the condition of possibility that allows us to gain knowledge about the world that surround us. ``Common sense'' plays here an important role securing the parameters of a ``common language'' which produces the illusion of an unproblematic (intersubjective) discourse about ``common observations''. From this viewpoint, the nature of observation should not be questioned. Accordingly, it is argued that if we begin by raw empirical data alone we are then starting by something ``pure'', ``uncontaminated'' and thus ``objective''.\footnote{Here the notion of objectivity is confused with that of intersubjectivity. A common mistake within philosophy of quantum mechanics since Bohr's account of physics.} From this perspective, the orthodox  philosophy of physics project focuses in trying to ``bridge the gap'' between our ``best (mathematical) theories'' and our common sense ``manifest image of the world''  \cite{Dorato15} ---an image derivative of the representation produced by classical physics in the XVII century. Following some of the main elements present within this very general line of thought, we might characterize then the epistemic view in terms of three main points. As the reader might recognize, each one of these points is quite commonly ---implicitly--- presupposed within many philosophical and foundational debates about quantum theory.

\begin{enumerate}
\item[{\bf Prediction of Measurement Outcomes:}] {\it Physical theories provide predictions about (``self  evident'') observable measurement outcomes and they do not necessarily provide a representation of physical reality.}
\item[{\bf Mathematical Formalism and Empirical Adequacy:}] {\it  A physical theory is a mathematical formalism which can be considered ---via a set of minimal interpretational rules--- as empirically adequate (or not).}
\item[{\bf Interpretations are Superfluous:}] {\it The interpretation of an empirically adequate theory is superfluous. It is a metaphysical exercise which cannot change the formalism of the theory nor the nature of the observations predicted by it.}
\end{enumerate}

Quantum mechanics has been characterized many times as one of ``our best physical theories''. It is empirically adequate and possesses a closed mathematical formalism. But after more than a century we do not possess ---up to the present--- any coherent representation of what the theory is really talking about. From the epistemic viewpoint this question might be regarded as an unimportant metaphysical enterprise which attempts to talk about something beyond observability. However, as we discussed above, the question of interpretation is not completely vanished even from epistemic and empiricist viewpoints.\footnote{The question of interpretation reappears in QM in different levels. See \cite{Cassini17}.} An empiricist such as van Fraassen might be interested for some reason or another in trying to find out what is the particular representation of physical reality provided by a particular interpretation (e.g., see van Fraassen's analysis of Rovelli's interpretation \cite{VF10}) even though he might not believe the theory to be true. 

Following our definitions, we might provide a general characterization of epistemic relationalism as a view that understands relations as derivative of observations, as a way to relate in a more coherent way such data. Once the data are observed, only then relations are introduced as a means of ``better resolving'' the strange problems which appear within the theory of quanta; this being done without actually questioning those fundamental ---conceptual--- presuppositions. Relations are not then taken as the basic elements out of which the world is made of, as elements of a systematic ontology capable of representing reality, but as an hypothesis added after the observations have been performed, in order ``to save'' in a ---maybe--- more precise way what we have already essentially supposed (i.e. classical phenomena). As we shall see, such observations might be characterized in different ways: in terms of `experimental arrangements', in terms of `facts', `perspectives' or even other `systems'. As we will now show, there are several interpretations of  quantum mechanics which have developed in different ways this particular understanding of (epistemic) relationalism.

\subsection{Bohr's Instrumental Relationalism}

Many elements present within the epistemic view we have just characterized might be associated to Bohr's pragmatic view of physics \cite{Bohr60} according to which: ``Physics is to be regarded not so much as the study of something a priori given, but rather as the development of methods of ordering and surveying human experience.'' Bohr ---in perfect line with the epistemic view--- considered quantum theory as an abstract symbolic formalism which had to be reduced through a limit (i.e., the correspondence principle) to classical physics and phenomena \cite{BokulichBokulich}. In this respect, Bohr also shared with the epistemic viewpoint ---even though in terms of a neo-Kantian perspective--- the idea that observable quantum phenomena are essentially ``classical phenomena''. According to the Danish physicist \cite[p. 7]{WZ}: ``[...] the unambiguous interpretation of any measurement must be essentially framed in terms of classical physical theories, and we may say that in this sense the language of Newton and Maxwell will remain the language of physicists for all time.'' In this respect [{\it Op. cit.}, p. 7], ``it would be a misconception to believe that the difficulties of the atomic theory may be evaded by eventually replacing the concepts of classical physics by new conceptual forms.'' Thus, taking distance from ontological and metaphysical problems, Bohr was maybe the first to develop an epistemic type of relationalism grounded on classical experimental situations. In his book, {\it The Philosophy of Quantum Mechanics}, Max Jammer discussed the attempt of Bohr to understand quantum mechanics is analogous fashion to relativity theory.

\begin{quotation}
\noindent {\small ``In 1929 Berliner decided to dedicate an issue of his journal to Max Planck in commemoration of the golden anniversary of his doctorate; he asked Sommerfeld, Rutherford, Schr\"odinger, Heisenberg, Jordan, Compton, London and Bohr to contribute papers and his request was answered in all cases. Bohr used this opportunity to expound in greater detail the epistemological background of his new interpretation of quantum mechanics. In his article he compared in three different aspects his approach with Einstein's theory of relativity. [...] Concerning the first two points of comparison Bohr was certainly right. But as to the third point of comparison, based on the assertion that relativity theory reveals `the subjective character of all concepts of classical physics' or, as Bohr declared again in the fall of 1929 in an address in Copenhagen, that `the theory of relativity remind us of the subjective character of all physical phenomena, a character which depends essentially upon the motion of the observer,' [...] Bohr overlooked that the theory of relativity is also a theory of invariants and that, above all, its notion of `events,' such as the collision of two particles, denotes something absolute, entirely independent of the reference frame of the observer and hence logically prior to the assignment of metrical attributes.'' \cite[p. 132]{Jammer74}}
\end{quotation} 

\noindent Jammer continues to say:

\begin{quotation}
\noindent {\small ``[...] in Bohr's relational theory, the question `What is the position (or momentum) of a certain particle' presupposes, to be meaningful, the reference to a specified physical arrangement [...] one may formulate a theory of `perspectives', the term perspective denoting a coordinated collection of measuring instruments either in the sense of reference systems as applied in relativity or in the sense of experimental arrangements as conceived by Bohr. The important point now is to understand that although a perspective may be occupied by an observer, it also exists without such an occupancy [...] A `relativistic frame of reference' may be regarded as a geometrical or rather kinematical perspective; Bohr's `experimental arrangement' is an instrumental perspective.'' \cite[p. 201]{Jammer74}}
\end{quotation}

\noindent Indeed, in relativity theory (like in classical mechanics) all events can be conceived as perfectly well defined events, meaning they can be always placed in a structure which allows us to think consistently of the actual existence of all present events. However, as expressed by the Kochen-Specker theorem \cite{KS}, this possibility is precluded in the orthodox quantum formalism. As we know, the multiple projection operators of a quantum state cannot be mapped to a global valuation of the Boolean elements $\{0,1\}$ (see for discussion de \cite{RFD14}).  

As one of us argued in \cite{deRonde15} Bohr might be regarded as responsible for introducing the linguistic turn into physics, confronting in this way the very naive conceptions of the praxis and original meaning of physics itself. Physics was then understood as being fundamentally grounded in language. Accordingly, {\it phúsis} and reality had to be considered only as words ---created by humans. A direct consequence of this development was that the ontological questions to which quantum mechanics was confronted in the first decades of the XX century had to be ``suspended''. Bohr's philosophy of physics played in this respect an essential role: ``We are suspended in language in such a way that we cannot say what is up and what is down. The word `reality' is also a word, a word which we must learn to use correctly.'' There is no quantum world but only a classical language in which we are trapped. As Wittgenstein had claimed: ``The limits of my language mean the limits of my world.'' Or, rephrasing it in Bohr's own terms: ``We must be clear that when it comes to atoms, language can be used only as in poetry. The poet, too, is not nearly so concerned with describing facts as with creating images and establishing mental connections.'' The Bohrian linguistic turn in physics was able to deconstruct reality through language. As a consequence, objectivity ---which presupposed a moment of unity related to an object--- became mere intersubjective agreement. Ontological questions in quantum mechanics were ---even more--- blurred, the relation between the experimental arrangement as described classically and that of which the mathematical formalism of quantum mechanics was talking about became then ``unspeakable''. 

According to Bohr's interpretation of quantum mechanics ---in analogous terms to positivism--- the (supposedly) objective character of the theory was secured by our classical language, a language which allowed us to refer to (classical) experimental apparatuses and phenomena:

\begin{quotation}
\noindent {\small ``On the lines of objective description, [I advocate using] the word {\it phenomenon} to refer only to observations obtained under circumstances whose description includes an account of the whole experimental arrangement.[...] The experimental conditions can be varied in many ways, but the point is that in each case we must be able to communicate to others what we have done and what we have learned, and that therefore the functioning of the measuring instruments must be described within the framework of classical physical ideas.'' \cite[p. 3]{WZ}}
\end{quotation}

\noindent Bohr wanted to bring together the multiple incompatible contexts through his own concept of complementarity. However, he was never able to answer the ontological questions which Einstein had posed to him once and again. He escaped the issue by always translating Einstein's ontological concerns into his own epistemological scheme of thought. But, as it is said, a translator is also a traitor. When Bohr's translation was finished ontology had been completely erased from the main discussion; once the job was done, he could then explain everything exclusively in terms of (classically described) experimental and measurement situations.\footnote{A particularly good example of Bohr's methodology can be found in his famous reply to Einstein Podolsky and Rosen \cite{Bohr35} where he also applied his complementarity principle and the idea that measurement situations define the representation of the state of affairs.} 

\subsection{Modal and Perspectival Relationalism}

During one of the famous conferences in Johensu organized by Kalervo Laurikainen in the eighties, Simon Kochen presented a relational type modal interpretation \cite{Kochen85}. The ideas presented there were regarded by Carl Friedrich von Weizs\"acker and Theodor Gornitz \cite{GW87} as ``an illuminating clarification of the mathematical structure of the theory, especially apt to describe the measuring process. We would however feel that it means not an alternative but a continuation to the Copenhagen interpretation (Bohr and, to some extent, Heisenberg).'' Kochen had proposed an ascription of properties based on the so called Schmidt theorem, inaugurating ---together with van Fraassen and Dieks--- what would become to be known, some years later, as ``modal interpretations'' (see \cite{Vermaas99a} for a detailed anlaysis). Within this framework, one is able to ascribe properties to the subsystems of a composite system in a pure state. The biorthogonal decomposition theorem (also called Schmidt theorem) is able to account for correlations between the quantum system and the apparatus considering the measurement and the actual observation as a special case of this representation. 

\begin{thm}
Given a state $|\Psi_{\alpha\beta}\rangle$ in $\cal H = \cal
H_{\alpha}\otimes \cal H_{\beta}$. The Schmidt theorem assures there
always exist orthonormal bases for $\cal H_{\alpha}$ and $\cal
H_{\beta}$, $\{|a_{i}\rangle\}$ and $\{|b_{j}\rangle\}$ such that
$|\Psi_{\alpha\beta}\rangle$ can be written as

\begin{center}$|\Psi_{\alpha\beta}\rangle = \sum c_{j}|a_{j}\rangle
\otimes |b_{j}\rangle$.\end{center}

\noindent The different values in $\{|c_{j}|^{2}\}$ represent the
spectrum of the state. Every $\lambda_{j}$ represents a projection
in $\cal H_{\alpha}$ and a projection in $\cal H_{\beta}$ defined as
$P_{\alpha}(\lambda_{j}) = \sum |a_{j}\rangle \langle a_{j}|$ and
$P_{\beta}(\lambda_{j}) = \sum |b_{j}\rangle \langle b_{j}|$,
respectively. Furthermore, if the $\{|c_{j}|^{2}\}$ are non
degenerate, there is a one-to-one correlation between the
projections $P_{\alpha} = \sum |a_{j}\rangle \langle a_{j}|$ and
$P_{\beta} = \sum |b_{j}\rangle \langle b_{j}|$ pertaining to
subsystems $\cal H_{\alpha}$ and $\cal H_{\beta}$ given by each
value of the spectrum.
\end{thm}

\noindent Through the Schmidt decomposition one can thus calculate
the states of the subsystems (which are one-to-one correlated)
obtaining:

\begin{equation}
\rho^{\alpha}=tr_{\beta}(|\Psi^{\alpha\beta}\rangle\langle\Psi^{\alpha\beta}|)=\sum_{i}
|c_{i}|^{2} |\alpha_{i}\rangle\langle\alpha_{i}|\end{equation}

\begin{equation}\rho^{\beta}=tr_{\alpha}(|\Psi^{\alpha\beta}\rangle\langle\Psi^{\alpha\beta}|)=\sum_{i}
|c_{i}|^{2} |\beta_{i}\rangle\langle\beta_{i}|
\end{equation}

\noindent These two states can be interpreted in a later stage as
representing the apparatus and the quantum system, respectively.
The different values ${|c_{i}|^{2}}$ represent the spectrum of the
Schmidt decomposition given by ${\lambda_{j}}$. Every
$\lambda_{j}$ represents a projection in $\cal H^{\alpha}$ and a
projection in $\cal H^{\beta}$ defined as $P^{\alpha}(\lambda_{j})=\sum|a^{\alpha}_{j}\rangle\langle
a^{\alpha}_{j}|$ and $P^{\beta}(\lambda_{j})=\sum|b^{\alpha}_{j}\rangle\langle
b^{\alpha}_{j}|$, respectively. Furthermore, if the ${|c_{i}|^{2}}$ are non degenerate,\footnote{In the
case of degeneracy it is also possible to define new multi-dimensional projections and recompose this one-to-one correlation between subsystems \cite{Dieks93}.} there is a one-to-one
correlation between the projections $P^{\alpha}(\lambda_{j})$ and $P^{\beta}(\lambda_{j})$ pertaining to subsystems $\alpha$ and $\beta$ given by each value of the spectrum $\lambda_{j}$. In other words, the state of a two-particle system picks out (in most cases, uniquely) a basis (and therefore an
observable) for each of the component systems. In this way, the projections of a two composite system $\alpha\beta$ defined from the Schmidt decomposition define the joint probability as:
\begin{equation}
p(P^{\alpha}(\lambda_{a}),P^{\beta}(\lambda_{b}))=\delta_{ab}
\end{equation}

\noindent So if $[P^{\alpha}(\lambda_{j})]=1$, which means that
the projection $P^{\alpha}(\lambda_{j})$ is certain, then
$[P^{\beta}(\lambda_{j})]=1$ with probability 1, and {\it vice
versa}. Kochen considers in particular the state
$\left|\Psi^{\alpha\beta}\right\rangle = \sum_{i} c_{i} \left|
a^{\alpha}_{i}\right\rangle \otimes |b^{\beta}_{i}\rangle$ that
one obtains after a von Neumann measurement, and interprets it in
terms of modalities: $\alpha$ {\it possibly possesses one of the
properties $| a^{\alpha}_{i}\rangle\langle a^{\alpha}_{i}|$, and
the actual possessed property $| a^{\alpha}_{k}\rangle\langle
a^{\alpha}_{k}|$ is determined by the observation that the device
$\beta$ possesses the reading $| b^{\beta}_{k}\rangle\langle
b^{\beta}_{k}|$}. In the biorthogonal decomposition there is thus,
a bi-univocal relation between the properties of the object and
the measuring device. So every pure state of a composite of two
disjoint systems should receive this interpretation. In this way
the {\it dynamical state}, with the use of the biorthogonal
decomposition, generates a probability measure over the set of
possible {\it value states}\footnote{The distinction between {\it
dynamical state} and {\it value state} was introduced by Bas Van
Fraassen in order to solve the inconsistencies into which one is
driven by the eigenstate-eigenvalue link. See
\cite{Dickson98, VF91}.}, namely the standard quantum
mechanical measure. As noted by Kochen (\cite{Kochen85}, p. 152): ``Every interaction gives rise to a unique correlation between certain canonically defined properties of the two interacting systems. These properties form a Boolean algebra
and so obey the laws of classical logic.'' Kochen's relational interpretation defines systems as ``being witnessed'' by one another:

\begin{quotation}
\noindent {\small ``In place of an official human observer, we assume that each system acts as witness to the state of the other... The world from this view becomes one of perspectives from different systems, with no privileged role for any one, and of properties which acquire a relational character by being realized only upon being witnessed by other systems.'' \cite[pp. 160-164]{Kochen85}}
\end{quotation}

The relation imposed by Kochen between systems that observe other different systems is of a type which resembles some of the views already discussed. It is interesting to notice that, as Vermaas [{\it Op. cit.}, p. 49] argues: ``If one accepts such relationalism; i.e. that properties are meaningful only with respect to a relative system; one can deny the need for correlations between the properties of all possible subsystems of a composite because, for Kochen, properties have a truly relational character.'' However, even though we agree with Vermaas' conclusion regarding the fundament of correlations within Kochen's approach, for our own purposes, at this point of our analysis it becomes of outmost importance to remark that: {\it relationalism is by no means equivalent nor implies necessarily relativism.} We might clarify this important point recalling the analysis we provided in the first part of this article. When we talked about sophistry, we analyzed Protagora's views in order to describe what the common ground of all relativisms is: {\it the reality of something is always relative to the individual perceiving subject who observes it.} ``Man is the measure of all things.'' This statement, as recognized by the Vienna Circle in their manifesto, is also close to empiricism which, in Locke's own wording, rephrased the Protagorian statement in terms of perception: ``To be, is to be perceived''. Both Bohr's and Kochen's proposals fall under this category. While for Bohr, quantum systems are defined in terms of (as being {\it relative} to) measurement situations, for Kochen the properties of systems are defined in terms of (as being {\it relative} to) other systems. As we shall see, the perspectival version of Bene and Dieks \cite{BeneDieks02}, and maybe, to some extent, even that of Rovelli, fall also under this general sophistic umbrella. We could say that they all propose ---in different ways--- extended Protagorisms: relativisms with broader definitions of what can be considered as an ``observer'' or ``perceiving subject''. 

Contrary to sophistry, in the sections above, we analyzed philosophical and physical theories that affirmed the possibility of representing {\it phúsis} beyond the dependence to actual {\it hic et nunc} observations. Some of them presented a world primarily made of independent substances; others, like Plato's {\it Sophist} and Spinoza's {\it Ethics}, denied ontological separation and proposed relation as fundamental within their own metaphysical representation of reality. For both Plato and Spinoza the world should be conceived as made of relations or potencies of relation. These ontological relational schemes expose the fact that relationalism does not entail relativism. Thus, at this point of our analysis, it is useful to distinguish between two different types of relations: \\

\noindent {\it {\bf Epistemic relations:} Relations as modeled from the empirical subject-object model. What is observed or perceived (the object or system) is not only related but also is ---more importantly--- relative to a subject (an agent, another system or an apparatus). Epistemic relations entail relativism, since we only know how things (observed or perceived) seem to be as relative to a given perspective, and never how they really are independently of a perceiving actor (an agent, another system or an apparatus).}\\ 

\noindent {\it {\bf Ontic relations:} Relations are the metaphysical building-blocks of reality, they are essential to the representation of phúsis. Relations exist within reality right from the start and their existence is absolutely independent of observations or a perceiving subject (an agent, another system or experimental arrangement).}\\   
 
Going now back to modal interpretations, we might remark that it is only in the case of {\it ontological}\footnote{Through this paper we use `ontic' and `ontological' as synonyms, in analogous manner to the way in which `epistemic' and `epistemological' are used in philosophy of physics.}  {\it relations} that it would make sense to look for joint probability distributions. This was, in fact, the main program in which Dieks, Bacciagaluppi, Clifton, Bub and other modal researchers engaged in at the beginning of the '90 (see \cite{DieksVermaas98}). However, in the case of {\it epistemic relations} it becomes meaningless to seek for such joint probability distribution. In such case, relations are intrinsically determined ---following Bohr's notion of contextuality \cite{deRonde16c}--- by the choice performed by a subject of the particular experimental set-up with which the object is studied. The problem is that due to the structure of the quantum formalism different choices of contexts determine incompatible local valuations which ---according to the Kochen-Specker theorem--- cannot be embedded into a whole global valuation \cite{deRonde16c}. For example, if we consider a composite system $\omega=\alpha\beta\gamma$ in a pure state and we take the two subsystems $\alpha\beta$ and $\alpha\gamma$, these two subsystems have properties which are
realized with respect to different relative systems. In general
the projections resulting from these subsystems will not be
commensurable and it will be not possible to obtain a classical
distribution for joint probabilities.\footnote{This result was explicitly addressed by Vermaas in his no-go theorem of 1997 \cite{Vermaas97}.} As remarked by Vermaas the impossibility to define joint property ascriptions does diminish ---at least from a realist viewpoint--- the attractiveness of such interpretations.

\begin{quotation}
\noindent {\small ``the spectral [K-D] modal interpretation is condemned to
perspectivalism. [...] If one accepts perspectivalism as discussed
[...] and as possibly embraced by Kochen (1985), the joint
probabilities give the correlations between all the properties one
can consider simultaneously. For, according to perspectivalism,
one can only simultaneously consider the properties of subsystems
if these systems can be considered from one and the same
perspective. [By] adopting perspectivalism most of the interesting
questions in quantum mechanics are simply evaded.'' \cite[p. 255]{Vermaas99a}}
\end{quotation}

\noindent Indeed, one considers in somewhat sophistic terms that ``the subject is the measure of all clicks'', the choice of the experimental arrangement appears then as a necessary precondition to account for what will be observed. Obviously, no individual subject or agent can adopt {\it hic et nunc} many different perspectives. An individual cannot be present, at one and the same time, in different viewpoints (contexts or frames of reference) which allow to observe a system. We, human subjects, perform observations always from our own particular perspective. In line with Bohr's proposal, the choice of the measurement arrangement must be then regarded as a precondition to define the system and its properties (see for a detailed discussion \cite{deRonde16c}). From this relativist stance, one can deny that Kochen's modal interpretation ever needs to say something sensible about the joint occurrence of the properties of $\alpha$ and $\beta$. The problem, as in sophistry, is that the consequence of this move erases any reference of the theory to a representation of physical reality (or {\it phúsis}) beyond observations or measurement outcomes. The connection between the theory and reality is then lost.

In 2002, a new perspectival version of the modal interpretation was developed by Gyula Bene and Dennis Dieks \cite{BeneDieks02} in which they continued the line of research proposed by Kochen.\footnote{For a detailed analysis of the Bene-Dieks perspectival interpretation see \cite{deRonde03}.} The central point of their interpretation can be summarized in the following passage: ``[...] instead of the usual treatment in which properties are supposed to correspond to monadic predicates, we will propose an analysis according to which properties have a relational character.'' Once again, relativism and relationalism are used as synonyms, assuming implicitly that relations must be necessarily understood as epistemic relations. As in the Kochen interpretation, in the Bene-Dieks (B-D for short) interpretation the state of a physical system {\it S} requires the specification of a ``reference system'' {\it R} with respect to which the state is defined. ``In this `perspectival' version of the modal interpretation properties of physical systems have a relational character and are defined with respect to another physical system that serves as a reference [or witnessing] system.'' 
\begin{equation}\rho_{S}(R)=Tr_{R/S}(\rho_{R}(R))\end{equation}

\noindent were $\rho_{S}(R)$ means that the system {\it S} is
witnessed from the perspective {\it R}; and $Tr_{R/S}$ means the
trace with respect to the degrees of freedom of {\it system R
`minus' S}. In the special case in which {\it R=S} the state  is
in general a one dimensional projector; i.e. the {\it state of S
with respect to itself}:
\begin{equation}\rho_{S}(S)=|\Psi_{S}\rangle\langle\Psi_{S}|\end{equation}

The state of {\it R w.r.t. itself}, $\rho_{R}(R)$, is postulated
to be one of the projectors contained in the spectral resolution
of $\rho_{R}(U/R)$, i.e. the state  $\rho_{R}$ from the
perspective of the {\it Universe `minus' system R}; and represents
the monadic properties of the system as in the K-D modal
interpretation. If there is no degeneracy  among the eigenvalues
of $\rho_{R}(U/R)$ these projectors are one dimensional and the
state can be represented by a vector $|\Psi_{S}\rangle$. The {\it
state of R w.r.t. itself} is given by one of the eigenvectors
$|R_{j}\rangle$ of $\rho_{R}(U/R) $. Kochen's witnessed state will
turn out to correspond to the state of the object with respect to
itself. The dynamical principle of the interpretation is that
$\rho_{U}(U)$ evolves unitarily in time. There is ``no collapse''
of the $|\Psi_{U}\rangle$ in this approach just as in ordinary
modal interpretations. The theory specifies only the probabilities
of the various possibilities and in this sense is an
indeterministic interpretation. Furthermore, it is assumed that
the state assigned to a closed system {\it S} undergoes a unitary
time evolution given by the Liouville equation:
\begin{equation}i\hbar\frac{\partial}{\partial t}\rho_{S}(S)=[H_{S},\rho_{S}(S)]
\end{equation}

\noindent If the systems $S_{1}, S_{2} ... S_{n}$ are pair-wise
disjoint and $U$ is the whole universe, then the joint probability
that $|\Psi_{S_{1}}\rangle$ coincides with
$|\varphi^{S_{1}}_{j_{i}}\rangle$, $|\Psi_{S_{1}}\rangle$
coincides with $|\varphi^{S_{1}}_{j_{i}}\rangle$, ...,
$|\Psi_{S_{1}}\rangle$ coincides with
$|\varphi^{S_{1}}_{j_{i}}\rangle$ is given by
\begin{equation}P(j_{1}, j_{2}, ..., j_{n}) = Tr(\rho_{U}(U) \prod_{i=1}^{n}
|\varphi^{S_{i}}_{j_{i}}\rangle\langle\varphi^{S_{i}}_{j_{i}}|)
\end{equation}

\noindent To summarize, because the existence of properties is always relative to a particular (perceiving) system, according to Bene and Dieks it makes no sense to compare properties from different perspectives: ``we do not define joint probabilities if the systems are not pair wise disjoint; in this way we block the no go probability theorem by Vermaas.'' Joint probabilities are only definable from a single definite {\it perspective R} and the restricted set of subsystems considered are thus commensurable ones.

\subsection{Rovelli's Informational Relationalism}

A similar analogy to that of Bohr between quantum mechanics and relativity was proposed by Carlo Rovelli in his now famous ``Relational Quantum Mechanics'' \cite{Rovelli96}. The notion explicitly rejected ---also, implicitly rejected in the interpretations of Bohr, Kochen and bene-diks--- is that of: absolute state or observer independent state of a system (observer-independent values of physical quantities). This notion is replaced in favor of: state relative to something. Rovelli argues that this necessity derives from the observation that the experimental evidence at the basis of quantum mechanics forces us to accept that distinct observers give different descriptions of the same events. There are two main ideas surrounding the interpretation of Rovelli, firstly, that the unease in quantum mechanics may derive from the use of a concept, which is inappropriate to describe the physical world at the quantum level; i.e. the notion of absolute state of a system. Secondly, that QM will cease to look puzzling only when we will be able to derive the formalism from a set of simple physical assertions (postulates, principles) about the world.

According to Rovelli we should derive the formalism from a set of experimentally motivated postulates just in the same way Einstein did for special relativity:

\begin{quotation}
\noindent {\small ``[...] Einstein's 1905 paper suddenly clarified the matter by pointing out the reason for the unease in taking Lorentz transformations `seriously': the implicit use of a concept (observer-independent time) inappropriate to describe reality when velocities are high. Equivalently: a common deep assumption about reality (simultaneity is observer-independent) which is physically untenable. The unease with the Lorentz transformations derived from a conceptual scheme in which an incorrect notion absolute simultaneity was assumed, yielding any sort of paradoxical consequences. Once this notion was removed the physical interpretation of the Lorentz transformations stood clear, and special relativity is now considered rather uncontroversial. Here I consider the hypothesis that all `paradoxical' situations associated with quantum mechanics as the famous and unfortunate half-dead Schr\"odinger cat [Schr\"odinger 1935] may derive from some analogous incorrect notion that we use in thinking about quantum mechanics. (Not in using quantum mechanics, since we seem to have learned to use it in a remarkably effective way.) The aim of this paper is to hunt for this incorrect notion, with the hope that by exposing it clearly to public contempt, we could free ourselves from the present unease with our best present theory of motion, and fully understand what does the theory assert about the world.'' [{\it Op. cit.}, p. 1639]}
\end{quotation}

Rovelli's interpretation takes distance from Bohr's distinction between macroscopic and microscopic systems. ``The disturbing aspect of Bohr's view is the inapplicability of quantum theory to macrophysics. This disturbing aspect vanishes, I believe, at the light of the discussion in this paper.'' Instead of the privileging certain observers (classical systems) Rovelli centers his interpretation in the concept of information. 

\begin{quotation}
\noindent {\small ``Information indicates the usual ascription of values to quantities that founds physics, but emphasizes their relational aspect. This ascription can be described within the theory itself, as information theoretical information, namely correlation. But such a description, in turn, is quantum mechanics and observer dependent, because a universal observer-independent description of the states of affairs of the world does not exist.'' [{\it Op. cit.}]}
\end{quotation}

Rovelli recognizes the impossibility of presenting an objective description in terms of systems and replaces this notion by ``net of relations''. According to him: ``[...] at the present level of experimental knowledge (hypothesis 2), we are forced to accept the result that there is no objective, or more precisely observer-independent meaning to the ascription of a property to a system. Thus, the properties of the systems are to be described by an interrelated net of observations and information collected from observations.'' [{\it Op. cit.}] The question becomes then: what can we say about this net of relations. Rovelli, talks about the notion of information: ``The notion of observer independent state of a system is replaced by the notion of information about a system that a physical system may possess.'' Still, as in the case of Bohr, Kochen, Bene and Dieks, the ontological question that any realist would want to answer is still present even though in a different form: information about what? Although it is possible to maintain a relational view of quantum states in terms of information, the ontological status of such information seems to remain a problematic issue ---at least, from a realist perspective.

\subsection{Revisiting Epistemic Relationalism: Relativism Reloaded?}

Our analysis has attempted to understand the common epistemological ground on which the just mentioned relational interpretations of quantum mechanics have been developed. As we have shown, all such interpretations take both a sophistic and empiricist standpoint. Sophistic in the sense that there is always the presupposition of someone or something (an agent, another system or even an apparatus) playing the role of a perceiving subject. Empiricist in the sense that observations are always considered as the ``self evident'' givens which allow us to produce knowledge. Even in the case of quantum measurements, the observed `clicks' in detectors are considered as being unproblematic ---as providing objective data. Such an epistemological (relativist) viewpoint conceives that knowledge is always of a perspectival nature, that it is fundamentally limited by observers, perceptions, other systems or even particular measurement situations. It also takes as a fundamental standpoint the idea that measurement outcomes or data can be ---in principle--- discussed without the need of applying a rigorous conceptual architecture. 

For our purposes, it becomes then important to remark that there is a common line of thought, a common agenda and set of presuppositions within the relational interpretations of quantum mechanics discussed above. While for Bohr the experimental arrangement is the measure of all (classical) phenomena, for Kochen it is always a system which acts as the measure (or witness) of another system. While Bene and Dieks argue ---following Bohr--- in favour of considering an instrumental perspective as a necessary condition of possibility to measure a quantum system, Rovelli seems to claim that all we have is the information that a system has about another system. Relationalism is then understood as implying relativism: the definition of a physical system and its properties (the object of study) is always relative to a witnessing subject, system or measurement apparatus (the subject which perceives). It is true that relativism implies always a relation between a subject and an object; but the opposite ---as we have already shown explicitly in previous sections--- is not true: relationalism is not necessarily committed to relativism. In fact, it is possible to understand relations in a completely non-relativist manner, namely, as ontological relations. To provide a clear understanding and definition of what is to be considered an ontic relation and how such ontological relationalism might help us to better understand the theory of quanta has been the main goal of this article.

\section{Ontological Relationalism in Quantum Mechanics: A New\\Proposal}

The ontic viewpoint ---as we understand it--- differs radically with respect to the epistemic account of physics. We can characterize the ontic view in terms of two main elements, firstly, the possibility of the theoretical representation of reality, and secondly, the denial of ``self evident'' or ``common sense'' observability. Returning to our introductory discussion, we might begin by stressing the positive characterization of the meaning of physics as a discipline which attempts to represent phúsis (or reality) in formal-conceptual terms. According to this viewpoint, physical theories provide, through the tight inter-relation of mathematical formalisms and networks of physical concepts, the possibility of representing both physical reality and experience. 

The possibility to imagine and picture reality beyond {\it hic et nunc} observation is provided not only by mathematical formalisms but also by physical concepts. Mathematics does not contain physical concepts, it does not represent anything beyond its own structure. One cannot derive as a theorem physical concepts from a mathematical system. Mathematicians can obviously work without learning about physical theories or the way in which physicists are able to relate formalisms with particular representations of physical reality. In fact, most mathematicians know nothing about physics and their work can be done without ever doing any type of experiment in a lab. A laboratory is completely useless for a mathematician. They neither require meta-physical concepts for their practice. The theory of calculus does not include the physical notions of Newtonian space and time, it does not talk about `particles', `mass' or `force'. In the same way, Maxwell's formalism cannot derive through a theorem the physical notion of `field'. Within physical theories, while mathematical formalisms are capable of providing a quantitative understanding, only conceptual schemes ---produced through the interrelation of many different concepts--- are capable of giving a qualitative understanding of physical reality and experience.

{\it Gedankenexperiments} are a good example of the power of conceptual and formal representations within physics. In fact, thought-experiments in physics have many times escaped the technical capabilities of their time and ventured themselves into debates about possible ---but unperformed--- experiences. Not only that, even impossible experiences ---such as those imagined by Leibinz and Newton regarding the existence of a single body in the Universe--- have been of great importance for the development of physics. Such physical ---possible or impossible--- counterfactual experiences can be only considered and imagined through an adequate conceptual scheme. Indeed, as remarked by Heisenberg \cite[p. 264]{Heis73}: ``The history of physics is not only a sequence of experimental discoveries and observations, followed by their mathematical description; it is also a history of concepts. For an understanding of the phenomena the first condition is the introduction of adequate concepts. Only with the help of correct concepts can we really know what has been observed.'' According to the ontic viewpoint, reality is not something ``self-evidently'' exposed through observations ---as positivists, empiricists and even Bohr has claimed---; on the contrary, its representation and understanding is only provided ---following the first physicists and philosophers--- through physical theories themselves. To avoid any misunderstanding, let us stress that the ontic viewpoint we are discussing here is not consistent with scientific realism, phenomenological realism or realism about observables all of which are in fact variants of empiricism grounded on common sense observability. As Einstein \cite[p. 175]{Dieks88a} made the point: ``[...] it is the purpose of theoretical physics to achieve understanding of physical reality which exists independently of the observer, and for which the distinction between `direct observable' and `not directly observable' has no ontological significance''. Observability is secondary even though ``the only decisive factor for the question whether or not to accept a particular physical theory is its empirical success.'' Empirical adequacy is part of a verification procedure, not that which ``needs to be saved'' ---as van Fraassen might argue\footnote{According to van Fraassen \cite[p. 197]{VF80}: ``the only believe involved in accepting a scientific theory is belief that it is empirically adequate: all that is {\it both} actual {\it and} observable finds a place in some model of the theory. So far as empirical adequacy is concerned, the theory would be just as good if
there existed nothing at all that was either unobservable or not actual. Acceptance of the theory does not commit us to belief in the
reality of either sort of thing.''}. Observability is something developed within each physical theory, it is a result of a theory rather than an obvious presupposition. At the opposite corner from the epistemic standpoint, Einstein \cite[p. 63]{Heis71} explained to Heisenberg that in fact: ``It is only the theory which decides what we can observe.'' Following these set of general considerations we might characterize what we have called elsewhere \cite{deRonde16b} representational realism:

\begin{enumerate}
\item[{\bf Physical Theory:}] {\it A physical theory is a mathematical formalism related to a set of physical concepts which only together are capable of providing a quantitative and qualitative understanding of a specific field of phenomena.}
\item[{\bf Formal-Conceptual Representation of Reality:}] {\it Physics attempts to provide theoretical, both formal and conceptual, representations of physical reality.}
\item[{\bf Observability is Created by the Theory:}] {\it The conditions of what is meant by `observability' are dependent on each specific theory. The understanding of observation is only possible through the development of adequate physical concepts.}
\end{enumerate}
   
To summarize, while the epistemic view considers that the world is accessible through ``common sense'' observation ---understood as a given---, which is also the key to develop scientific knowledge itself; the ontic viewpoint takes the opposite standpoint and argues that it is only through the creation of theories that we are capable of providing understanding of our experience in the world. According to the latter view, the physical explanation of our experience goes very much against ``common sense'' observability. The history of physics can be also regarded as the continuous change in our ``common sense'' understanding of the world. It was not obvious for the contemporaries of Newton that the same force commands the movement of the moon, the planets and a falling apple. It was not inescapable in the 18th Century that the strange phenomena of magnetism and electricity could be unified through the strange notion of electromagnetic field. And it was far from evident ---before Einstein--- that space and time are entangled, that objects shrink and time dilate with speed. To sum up, we maintain that what is needed is a conceptual representation of what the quantum formalism is expressing, and not merely a salvaging of the relation between our ``common sense'' understanding of reality and measurement outcomes ---sweeping under the classical carpet the most interesting, effective and productive aspects of the formalism. What QM talks about ---we argue--- seems difficult to be grasped through a substantivalist (atomistic) understanding of reality that supposes individual separated substances. After more than a century trying to fit the quantum formalism into such a presupposed metaphysical representation of reality it might be time to try something new. 

Our proposal is to develop, taking inspiration from some of the elements found in the revisions of both Plato's and Spinoza's philosophies, a truly relational ontology (this is, one that considers relation as being fundamental) which is capable of providing a new (representational) realist way of understanding the theory of quanta. Both philosophers' understanding of `potency' or `possibility' in ontological terms, as well as the connection between that understanding and their relational views ---which, as we saw, are capable of articulating a specific knowledge of the world without producing substantial separations---, might allow us to throw new light on some key features of the quantum formalism such as: contextuality, superposition, non-individuality, non-separability, etc. The specific consideration of these features in ontic relational terms will be addressed in future works.

\section*{Acknowledgements} 

This work was partially supported by the following grants: FWO project G.0405.08 and FWO-research community W0.030.06. CONICET RES. 4541-12 and the Project PIO-CONICET-UNAJ (15520150100008CO) ``Quantum Superpositions in Quantum Information Processing''.


\end{document}